\pdfoutput=1	%
\documentclass[10pt,twocolumn,twoside]{pnas-new}
\setboolean{displaywatermark}{false}

\templatetype{pnasresearcharticle} %

\title{Synthesizing Compact Hardware for Accelerating Inference from Physical Signals in Sensors}

\author[1]{Vasileios Tsoutsouras}
\author[2]{Max Vigdorchik}
\author[3]{Phillip Stanley-Marbell} 

\affil[1,2,3]{Department of Engineering, University of Cambridge, Cambridge CB3 0FA, UK.}

\leadauthor{Tsoutsouras} 

\authorcontributions{PSM and VT conceived the idea. VT performed the compiler RTL-generation backend implementation. VT and MV performed the experimental evaluation. All authors contributed to the writing.}
\correspondingauthor{\textsuperscript{1}To whom correspondence should be addressed. E-mail: {vt298@cam.ac.uk}}

\vspace{-0.1in}

\usepackage{framed}
\usepackage{moreverb}
\usepackage{xspace}
\usepackage{flushend}			%

\usepackage{xfrac}			%

\usepackage{amsfonts}
\usepackage{amssymb}    
\usepackage{amsmath}
\usepackage[thmmarks,amsmath,hyperref]{ntheorem}
\usepackage{graphicx}
\usepackage{times}
\usepackage[scaled]{helvet} 		%
\usepackage{textcomp}
\usepackage{color}
\usepackage{booktabs}
\usepackage{listings}
\usepackage{multicol}
\usepackage{times}
\usepackage{pifont}			%
\usepackage{microtype}

\usepackage{mmm}
\usepackage{wand-sample}
\usepackage[scaled=0.85]{beramono}
\usepackage[T1]{fontenc}
\usepackage{url}
\usepackage{multirow}
\usepackage{array}

\newcommand\code[1]             {{\bf\ttfamily\small #1}}

\definecolor{listinggreen}{rgb}{0,0.6,0}
\definecolor{listinggray}{rgb}{0.5,0.5,0.5}
\definecolor{listingmauve}{rgb}{0.58,0,0.82}
\definecolor{listingkeywordcolor}{rgb}{1.0,0.4,0.0}
\definecolor{listinglightgray}{rgb}{0.9863,0.9863,0.9863}

\lstdefinelanguage{FSharp}
{
	morekeywords	= {
    let,
    type,
    Measure,
	},
	sensitive	= false,
	morecomment	= [l]{\#},
	morecomment	= [s]{(*}{*)},
}

\lstdefinelanguage{Newton}
{
	morekeywords	= {
		signal,
		derivation,
		symbol,
		name,
		invariant,
		constant,
		English,
		sensor,
		name,
		none,
		dot,
		cross,
		derivative,
		integral,
		interface,
		i2c,
		spi,
		analog,
		write,
		read,
		delay,
		range,
		erasuretoken,
		uncertainty,
		accuracy,
		precision,
		Gaussian,
		exponential,
		biexponential,
		to,
		bits,
		dimensionless,
		include
	},
	sensitive	= false,
	morecomment	= [l]{\#},
	morecomment	= [s]{/*}{*/},
}

\usepackage{hyperref}
\hypersetup{
    colorlinks=false, %
    linktoc=all,     %
    linkcolor=blue,  %
}

\begin{abstract}
We present \emph{dimensional circuit synthesis}, a new method for
generating digital logic circuits that improve the efficiency of
training and inference of machine learning models from sensor data.
The hardware accelerators that the method generates are compact
enough (a few thousand gates) to allow integration within low-cost
miniaturized sensor integrated circuits, right next to the sensor
transducer.  The method takes as input a description of physical
properties of relevant signals in the sensor transduction process
and generates as output a Verilog register transfer level (RTL)
description for a circuit that computes low-level features that
exploit the units of measure of the signals in the system.

\qquad We implement dimensional circuit synthesis as a backend to
the compiler for Newton, a language for describing physical systems.
We evaluate the backend implementation and the hardware it
generates, on descriptions of 7 physical systems.  The results show
that our implementation of dimensional circuit synthesis generates
circuits of as little as 1662 logic cells / 1239 gates for the systems
we evaluate.

\qquad We synthesize the designs generated by the dimensional circuit
synthesis compilation backend for a low-power miniature FPGA targeted
by its manufacturer at sensor interface applications. The circuits
which the method generated use as little as 27\% of the resources
of the 2.15$\times$2.5\,mm FPGA. We measure the power
dissipation of the FPGA's isolated core supply rail and show that,
driven with a pseudorandom signal input stream, the synthesized
designs use as little as 1.0\,mW and no more than 5.8\,mW. These
results show the feasibility of integrating physics-inspired machine
learning methods within low-cost miniaturized sensor integrated
circuits, right next to the sensor transducer.
\end{abstract}

\dates{This manuscript was compiled on \today}

\definecolor{a}{rgb}{0.9,0.95,0.95}%
\definecolor{b}{rgb}{0.99,0.99,0.99}%

\begin{document}

\verticaladjustment{-5pt}

\maketitle
\thispagestyle{firststyle}
\ifthenelse{\boolean{shortarticle}}{\ifthenelse{\boolean{singlecolumn}}{\abscontentformatted}{\abscontent}}{}

\section{Introduction}
Sensor integrated circuits are at the forefront of the data pipeline
feeding the recent revolution in machine learning systems. Sensors
transduce a physical signal such as acceleration, temperature, or
light, into a voltage which is then converted by analog-to-digital
converters (ADCs) into a numeric representation, for input to
computation. Digital preprocessing within sensor integrated circuits,
or the software that consume their output, then apply appropriate
calibration constants and scaling to convert these digitized voltages
into a scaled and dimensionally-meaningful representation of the
signal (e.g., acceleration in $m/s^2$).

\begin{figure}
\centering
\includegraphics[trim=0cm 0cm 0cm 0cm, clip=true, angle=0, width=0.485\textwidth]{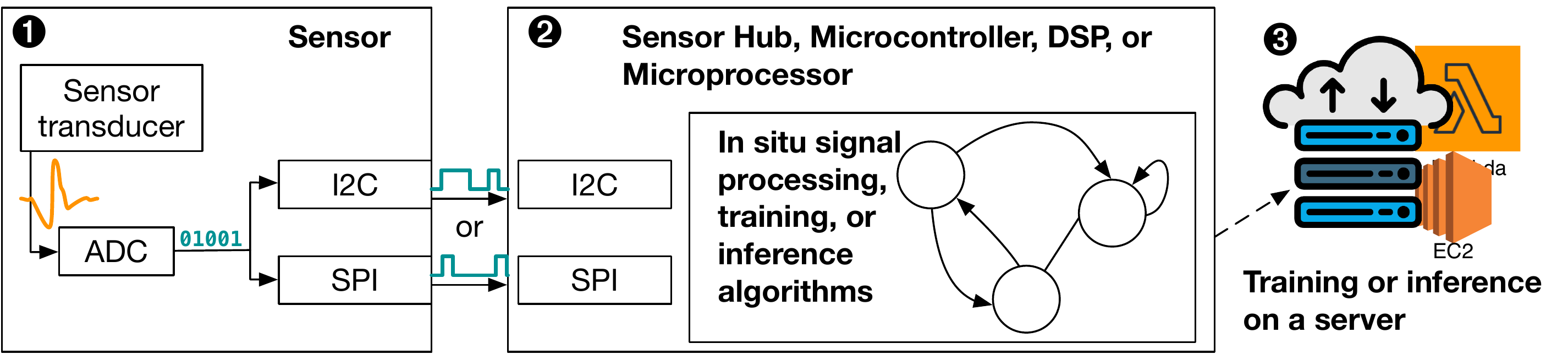}
\caption{Existing sensing systems typically send data to servers
for training models and generating inferences.  Moving data both
within a system and over networks adds latency and costs energy.}
\label{fig:introduction:sensors-in-systems}
\end{figure}

Figure~\ref{fig:introduction:sensors-in-systems} shows how contemporary
sensor-driven computing systems move the digitized data at the
output of signal conversion circuits through many transmission and
storage steps before the data are used in training a model or in
driving an inference, typically on server far removed from the
sensing process. This data movement costs time and energy.  When
ever-greater volumes of data will enable new applications and
inference models, it will be valuable to perform the necessary
computations as close to the signal acquisition and transduction
process as possible: ideally, in the sensor integrated circuit
itself (labeled ``\ding{202}'' in
Figure~\ref{fig:introduction:sensors-in-systems}).

However, since these sensor integrated circuits are typically
required be low cost (often under 10\,USD), have small die area
(often less than 4\,mm$^2$), and use minimal power (typically under
1\,mW), it is challenging to integrate even the most efficient and
compact traditional learning and inference methods into these devices
themselves.

\subsection{Physics constrains signals from sensors}
The values taken on by data from sensors are constrained by the
laws of physics and by the dynamics of the structures to which
sensors are attached. Most physical laws and the governing equations
for most system dynamics take the form of sums of product terms
with each product term comprising powers of the system's
variables~\cite{feynman1967character, feynman1965feynman,
Buckingham1914}.  Because there are a bounded number of irreducible
relative powers in these product terms whose units of measure result
in meaningful units for the whole expression, it is common in many
engineering disciplines to use information on units of measure
(\textit{dimensional analysis}) to derive candidate relations for
experimentally-observed phenomena~\cite{mahajan2010street,
Buckingham1914}. Recent work~\cite{10.1145/3358218} has used this
observation to prune the hypothesis set of functions considered
during machine learning to achieve significant improvements in both
training and inference, improving training latency by 8660$\times$
and reducing the arithmetic operations in inference over 34$\times$.

In this work, we build on these results to develop a new backend
for the Newton compiler~\cite{lim2018newton}. The backend generates
register transfer level (RTL) hardware designs for accelerating the
execution of the required pre-inference parts of analytic models
relating the signals in a multi-sensor system according to the
specifications of the physics of a system. Because our method uses
information from dimensional analysis and units of measure to
synthesize hardware to accelerate inference from sensors, we call
the method \emph{dimensional circuit synthesis}.

\subsection{Dimensional circuit synthesis: physics-derived pre-inference processing in sensors}
Dimensional circuit synthesis is a compile-time method to generate
digital logic circuits for performing pre-inference processing on
sensor signals. Dimensional circuit synthesis takes as its input a
specification of the signals that can be obtained from the sensors
in a system and their units of measure.  Using these specifications,
dimensional circuit synthesis generates hardware to compute a set
of physically plausible expressions relating the signals in the
system.  The logic circuits which the method generates represent
sets of monomial expressions which form dimensionless groupings of
sensor signals (i.e., products whose units cancel out).  The
synthesized hardware takes as input digital representations of
sensor readings and generates the computed value of the dimensionless
expressions as its output. A machine learning training or inference
process then uses these dimensionless products as its inputs and
prior work has shown that this preprocessing can significantly
improve both the latency and accuracy of inference. An on-device
(in-sensor) inference engine will integrate the generated RTL that
performs pre-processing with either custom RTL or a programmable
core implementing the inference using, e.g., a neural network.
We evaluate the generated RTL on the Lattice iCE40, a
state-of-the-art, ultra-miniature FPGA that meets the size
and power consumption constraints of in-sensor processing.

\subsection{Contributions}
This article makes two main contributions to on-device and in-sensor 
inference:
\begin{itemize}
\item We present dimensional circuit synthesis
(Section~\ref{sec:methodology}), a new method to generate RTL
hardware for pre-processing sensor data prior to inference, thereby
improving latency and reducing overhead.

\item We evaluate the generated RTL on the Lattice
iCE40 ultra-miniature FPGA (Section~\ref{sec:results}) and show
that the generated RTL is fast enough to allow real-time processing,
while consuming minimal power.
\end{itemize}

\begin{figure}
\begin{lstlisting}[language=Newton]
include "NewtonBaseSignals.nt"

v0	: constant = 0 (meter*second**-1);

UAVglider: invariant(
					h: distance,
					v: speed,
					m: mass) =
{
	h ~ {v, m, v0, kNewtonUnithave_AccelerationDueToGravity}
}
\end{lstlisting}
\caption{Newton language specification, for a
sensor-instrumented unpowered glider. The specification only states
the physical signals/quantities relevant to the system, their units
of measure, and the fact that they are related to each other and
to a constant \code{kNewtonUnithave\_AccelerationDueToGravity} (line
10).}
\label{figure:methodology:NewtonExample}
\end{figure}

\section{Background and Methodology}
\label{sec:methodology}
\label{subsec:dfs}
\begin{figure}
\includegraphics[trim=0cm 0cm 0cm 0cm, clip=true, angle=0, width=1.0\columnwidth]{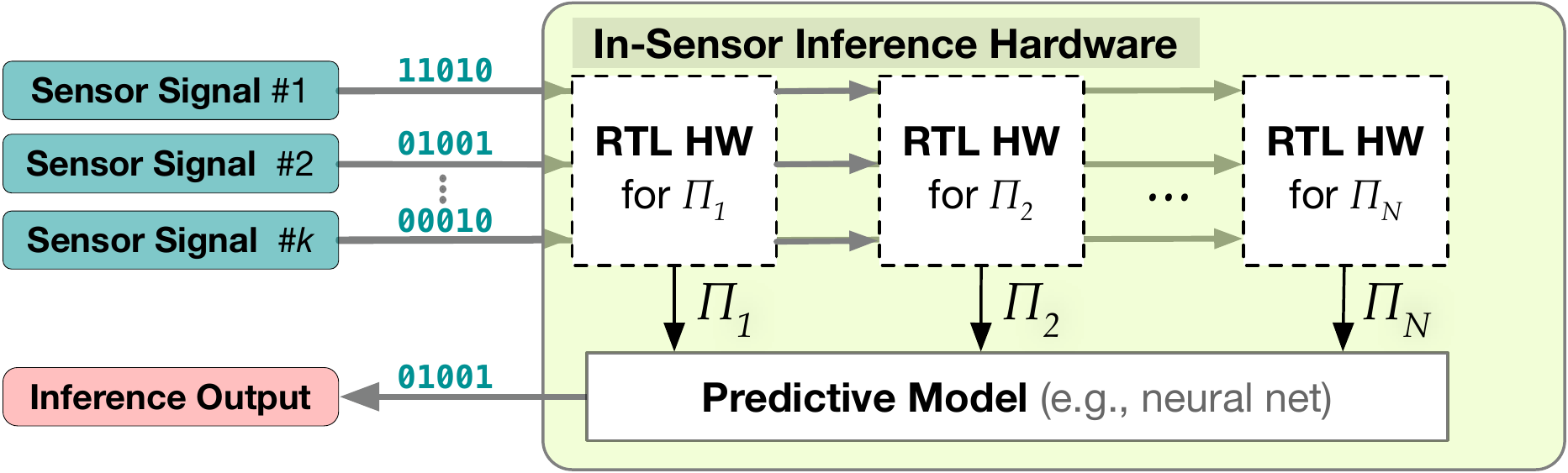}
\caption{The hardware generated by dimensional circuit synthesis
preprocesses $k$ sensor signals to obtain $N < k$ \emph{dimensionless
products} $\Pi_1 \ldots \Pi_k$.  A predictive model takes these
products as input and generates an inference output.}
\label{fig:introduction:dfsRTL}
\end{figure}

Dimensional circuit synthesis takes as input descriptions of the
units of measure of the sensor signals in a system.
Figure~\ref{figure:methodology:NewtonExample} shows an example
Newton description for an unpowered UAV (i.e., a glider).  Let a
physical system for which we want to construct an efficient predictive
model have $k$ symbols corresponding to physical constants or sensor
signals.  From the Buckingham $\Pi$-theorem~\cite{Buckingham1914},
we can form $N \le k$ dimensionless products, $\Pi_1 \ldots \Pi_N$
and these dimensionless products are the roots of some function
$\Phi$, where
\begin{align}
\label{equation:Phi}
\Phi (\Pi_1, \Pi_2, \dots, \Pi_i, \ldots, \Pi_{N}) = 0 .
\end{align}
Wang \textit{et al.}~\citep{10.1145/3358218} use
Equation~\ref{equation:Phi} as the basis for generating a preprocessing
step of offline training and inference of models of physical systems
and propose an automated framework for generating these dimensionless
products.  In a subsequent calibration step, they learn a model for
the function $\Phi$ and demonstrate that learning $\Phi$ from the
$\Pi_1 \ldots \Pi_N$ can be both significantly more efficient and
more accurate than learning a function from the original $k$ sensor
signals directly. The dimensionless products $\Pi_1 \ldots \Pi_N$
are essential to both training and inference and to achieving the
orders-of-magnitude speedup.  In this work, we present  a method
for generating hardware to efficiently compute these $\Pi$s. Doing
so close to the sensor transducer also reduces the data sensing
systems must transmit from the sensor transducer to either a sensor
hub, microcontroller, or other component performing on-device
training and inference, potentially improving system efficiency and
performance.  Figure~\ref{fig:introduction:dfsRTL} shows how the
generated hardware for $\Pi$ computation fits within an on-device
inference system.

\subsection{Dimensional Circuit Synthesis}
Figure~\ref{fig:introduction:dfsRTL} shows how hardware blocks
generated by dimensional circuit synthesis calculate the values of
the $\Pi$ products.  The input of these modules are the sensor
signals corresponding to the physical parameters specified as the
input to the dimensional circuit synthesis analysis in the Newton
specification language (see, e.g.,
Figure~\ref{figure:methodology:NewtonExample}).  The calculated
$\Pi$ product values correspond to the output of the pre-processing
step of the inference function and they feed into any existing
method for classification or regression.  This final step could be
a programmable low-power core such as the 32-bit RISC cores now
integrated into some state-of-the-art sensor integrated circuits
or a low-power machine learning accelerator such as
Marlann~\cite{SymbioticEDA:Marlann}, implemented in either RTL or
in a miniature FPGA like we use in our evaluation in
Section~\ref{sec:results}.

\begin{figure*}
\centering
\includegraphics[trim=0cm 0cm 0cm 0cm, clip=true, angle=0, width=2.0\columnwidth]{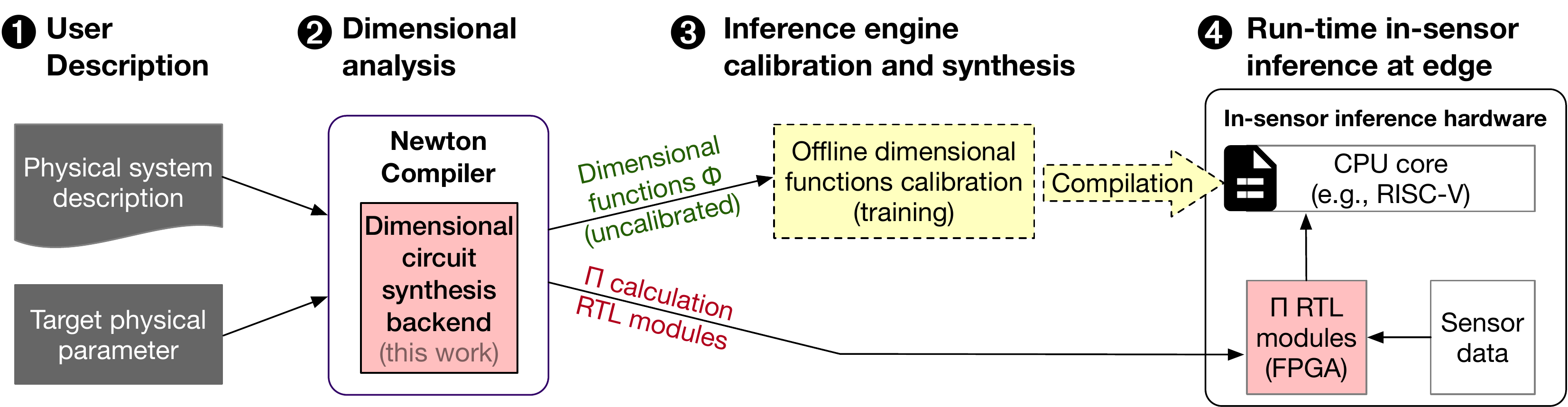}
\caption{Proposed dimensional circuit synthesis framework. In
Step~\ding{202}, the users provide the specifications and target
inference parameter of the examined physical system. Newton compiler
including our implemented backend is executed in Step~\ding{203}.
In Step~\ding{204} the framework translates the generated RTL modules
to an FPGA bitstream and in parallel we execute a manual calibration
of dimensional functions~\cite{10.1145/3358218} (box with dashed
border).  In Step~\ding{205} the generated SW and HW modules are
downloaded to the in-sensor inference engine.}
\label{fig:introduction:dfsCompilationProcess}
\end{figure*}

\subsubsection{Approach and implementation}
Figure~\ref{fig:introduction:dfsCompilationProcess} shows the four
steps which make up our implementation of dimensional circuit
synthesis.  We implemented these steps as a new backend of the
Newton compiler~\cite{lim2018newton}, but the techniques in
Figure~\ref{fig:introduction:dfsCompilationProcess} could in principle
be applied to any specification for physical systems that contains
information on units of measure.

In Step~\ding{202}, a user of dimensional circuit synthesis creates
a Newton language description such at that in
Figure~\ref{figure:methodology:NewtonExample}, specifying the
physical signals that describe the target physical system and from
which a machine learning model will eventually be trained.

Next, in Step~\ding{203}, the user 
invokes the Newton compiler with our new dimensional circuit synthesis
backend activated.  Because the method of constructing dimensionless
groups can result in multiple dimensionless products ($\Pi_i$ in
Equation~\ref{equation:Phi}), the user specifies which of the physical
signals in the input physical system description will be the target
variable of a machine learning model for the function $\Phi$ from
Equation~\ref{equation:Phi}.  Our new dimensional circuit synthesis
backend identifies the group of dimensionless products where the
target parameter appears in only one of the dimensionless products.
The outputs of Step~\ding{203} are: (i) a function $\Phi$, defined
in terms of the dimensionless products $\Pi_i$, but whose form has
not yet been fully defined; (ii) RTL descriptions for hardware to
compute the dimensionless products $\Pi_i$, including RTL descriptions
of the functional units (multipliers and dividers) that will perform
the arithmetic operations of the dimensionless product monomials.
Because floating-point operations can be expensive in both resources
and execution latency on energy-constrained on-device training and
inference systems, we use a signed fixed-point approximate real
number representation~\cite{behrooz2000computer} for the signals
in the dimensionless products computed in the synthesized hardware.
Each real number is represented by 32 bits, using 1 bit for the
sign, 16 bits for the decimal part and 15 bits for the fractional
part (i.e., a Q16.15 fixed-point representation).  This choice leads
to fast and lightweight multiplication and division units by
sacrificing the ability to use an arbitrary precision floating point
representation. The compiler backend is fully parametric with respect
to the length of the fixed point representation as well the precision
of the fractional part and can generate hardware with arbitrary
fixed-point representation sizes. This will allow future designs
to tailor the precision of the compute modules to the requirements
of the inference algorithms~\cite{micikevicius2017mixed}.

In Step~\ding{204}, we can train the uncalibrated dimensional
function offline on values of the dimensionless groups $\Pi_i$
computed offline as done in prior work~\cite{10.1145/3358218}.

Finally, in Step~\ding{205} the outputs of the hardware blocks
computing the dimensionless products feed the models trained offline
to generate inferences.  Alternatively a system could also use the
values of the dimensionless products to feed in situ training of
models implemented in a processor core, or to feed training in situ
of a hardware neural network accelerator~\cite{SymbioticEDA:Marlann}.

\section{Experimental Evaluation}
\label{sec:results}
\label{subsec:experimental-setup}
We evaluated the hardware generated by the dimensional circuit
synthesis backend using a Lattice Semiconductor iCE40 FPGA. The
iCE40 is a low-power miniature FPGA in a miniature wafer-scale
2.15\,mm$\times$2.50\,mm WLCSP package and is targeted at sensor
interfacing tasks and at on-device machine learning.  We used the
fully open-source FPGA design flow, comprising the
YoSys~\cite{shah2019yosys+} synthesis tool (version 0.8+456) for
synthesis and NextPNR ~\cite{shah2019yosys+} (version git sha1
5344bc3) for placing, routing, and timing analysis.

We performed our measurements on an iCE40 Mobile Development Kit (MDK)
which includes a 1$\Omega$ sense resistor in series with each of
the supply rails of the FPGA (core, PLL, I/O banks).  We measure the current
drawn by the FPGA core by measuring the voltage drop across the
FPGA core supply rail (1.2\,V) resistor using a Keithley DM7510, a
laboratory-grade 7-\sfrac{1}{2} digital multimeter that can measure
voltages down to 10\,nV and we thereby computed the power dissipated
by the FPGA core for each configured RTL design.
We used a pseudorandom number generator to feed the $\Pi$ computation
circuit modules under evaluation with random input data.

\begin{table*}
        \centering
        \caption{Experimental evaluation on iCE40 FPGA of dimensional circuit modules generated from the description of 7 physical systems.} \label{table:results}
        \tiny
        \begin{tabular}{lm{3.0cm}lllllll}
                \toprule
                \textbf{Name}			& \textbf{Description}	& \textbf{Target}		&\textbf{LUT4}		&\textbf{Gate}	&\textbf{Maximum}	&\textbf{Execution}	&\textbf{Avg. Power}		&\textbf{Avg. Power}\\
                                       &						& \textbf{Parameter}	&\textbf{Cells} 	&\textbf{Count}	&\textbf{Frequency}	&\textbf{Latency}	&\textbf{at 12 MHz}		&\textbf{at 6 MHz}\\
                \hline
\textbf{Beam}					& Cantilevered beam model, excluding mass of beam						& Beam deflection	& 2958 & 2590 & 16.88\,Mhz & 115 cycles & 3.5\,mW & 1.8\,mW\\
\textbf{Pendulum, static} 		& Simple pendulum excluding dynamics and friction						& Osc. period 		& 1402 & 1239 & 17.07\,Mhz & 115 cycles & 2.0\,mW & 1.1\,mW\\
\textbf{Fluid in Pipe} 			& Pressure drop of a fluid through a pipe								& Fluid velocity 	& 4258 & 3752 & 15.65\,Mhz & 188 cycles & 5.8\,mW & 3.0\,mW\\
\textbf{Unpowered flight} 		& Unpowered flight (e.g., catapulted drone)							& Position (height)	& 1930 & 1865 & 16.44\,Mhz & 81 cycles & 2.3\,mW & 1.2\,mW\\
\textbf{Vibrating string}		& Vibrating string														& Osc. frequency 	& 2183 & 1787 & 16.67\,Mhz & 183 cycles & 2.5\,mW & 1.3\,mW\\
\textbf{Warm vibrating string}	& Vibrating string	with temperature dependence							& Osc. frequency 	& 3137 & 2718 & 16.77\,Mhz & 269 cycles & 1.9\,mW & 1.0\,mW\\
\textbf{Spring-mass system}		& Vertical spring with attached mass 									& Spring constant 	& 1419 & 1240 & 16.67\,Mhz & 115 cycles & 3.4\,mW & 1.8\,mW\\
                \bottomrule
        \end{tabular}
\end{table*}

\subsection{Results}
We evaluated dimensional circuit synthesis on seven different
physical systems described in the Newton specification language.
Table~\ref{table:results} provides a brief description of the inputs
as well as a summary of the measurement results.  The table also
includes the target parameter for each respective execution of the
Newton compiler.  For example, for the physical description of the
pendulum the target parameter was its oscillation period, while for
the physical descriptions of a fluid in pipe the target parameter
was the velocity of the fluid.  This value of this parameter is
inferred at run-time by the machine learning model that is fed with
the output of the $\Pi$ computation.  The results in
Table~\ref{table:results} show the FPGA resource utilization as
well as resource utilization when mapped to CMOS gates, of each
generated $\Pi$ computation module, including the fixed point
arithmetic modules that implement the required arithmetic operations.

The execution latency column lists the cycles required for
completing the calculation of each of the generated RTL modules.
We obtained the number of cycles by simulating the execution of the
RTL modules for pseudorandom inputs generated by an LFSR. In each
RTL module, the calculation of different $\Pi$ products is parallelized
but the required operations per $\Pi$ product are executed serially.
As a result, designs with larger resource usage in the table, such
as the hardware for the unpowered flight model, conclude faster
than smaller designs such as the static pendulum. All modules require
less than 300 cycles. As a result, for both 6 and 12\,Mhz clocks,
the generated hardware can handle sample rates of over 10k
samples/second, permitting real-time operation.

The last column of Table~\ref{table:results} shows the measured
power dissipation of each design running in the iCE40 FPGA. In all
cases, the power dissipation is less than 6\,mW and as low as 1\,mW,
demonstrating the suitability of our method for small-factor,
battery-operated on-device inference at the edge.

\section{Insights and Conclusions}
\label{section:conclusions}
\emph{Dimensional circuit synthesis} is a new method for generating
digital logic circuits that improve the efficiency of training and
inference of machine learning models from sensor data. The method
complements prior work on\emph{ dimensional function synthesis}, a
new method for learning models from sensor data that enables orders
of magnitude improvements in training and inference on physics-constrained
signal data.  Dimensional circuit synthesis, which we present in
this paper, implements preprocessing steps required by dimensional
function synthesis in hardware.  This article presented the principle
behind the methods and the design and implementation of a compiler
backend that implements dimensional circuit synthesis for the iCE40,
a low-power miniature FPGA in a miniature wafer-scale
2.15\,mm$\times$2.50\,mm WLCSP package which is targeted at sensor
interfacing tasks and at on-device machine learning.  The hardware
accelerators that the method generates are compact (fewer than four
thousand gates for all the examples investigated) and low power
(dissipating less than 6\,mW even on a non-optimum FPGA).  These
results show, for the first time, that it could be feasible to
integrate physics-inspired machine learning methods within low-cost
miniaturized sensor integrated circuits, right next to the sensor
transducer.

\vspace{0.25in}

\bibliography{DFS-mlsys-2020-submission}

\onecolumn
\appendix

\twocolumn

\end{document}